\documentclass[letterpaper,12pt]{article}   %% LaTeX 2e (preferred)
\usepackage{osajnl2} %% do not use with REVTeX4
\usepackage[draft]{hyperref} %% optional

\begin{document}

\title{A frequency doubled 1534 nm laser system for potassium laser cooling.}

%% For REVTeX it is possible to automate superscript and e-mail callouts with the superscriptaddress option; see REVTeX4 documentation.

\author{Guillaume Stern,$^{*}$ Baptiste Allard, Martin Robert-de-Saint-Vincent,  Jean-Philippe Brantut, Baptiste Battelier, Thomas Bourdel, and Philippe Bouyer.}

\address{Laboratoire Charles Fabry de l'Institut d'Optique, \\Centre National de la Recherche Scientifique et Universit\'{e} Paris Sud 11, \\Institut d'Optique Graduate School, RD 128,\\ 91127 Palaiseau Cedex, France}
\address{$^*$Corresponding author: guillaume.stern@institutoptique.fr}

\begin{abstract}We demonstrate a compact laser source suitable for the trapping and cooling of potassium. By frequency doubling a fiber laser diode at 1534 nm in a waveguide, we produce 767 nm laser light. A current modulation of the diode allows to generate the two required frequencies for cooling in a simple and robust apparatus. We successfully used this laser source to trap $^{39}$K.\end{abstract}

% REPLACE WITH CORRECT OCIS CODES FOR YOUR ARTICLE
                          % NOTE: \ocis{} IS ALIASED TO \pacs{} BUT MUST
                          % FORMAT THE TERMS CORRECTLY FOR EACH JOURNAL

\maketitle %% null function with osajnl.sty

\section{Introduction}

Nowadays, different applications, such as atomic clocks or atomic inertial sensors, require simple and transportable systems \cite{Konemann07}, especially for space-based projects \cite{Laurent06}. Laboratory experiments on cold atoms, which are more and more complex, would also greatly benefit from simplification of the laser cooling setups. These laser systems need to combine narrow linewidths and powers of few hundreds of mW.

The most common laser source for alkaline cooling is the use of semiconductor laser diodes in an external cavity eventually further amplified with slave diodes or semiconductor tapered amplifiers. Recently, new solutions relying on telecoms technologies and second harmonic generation (SHG) \cite{Thompson03, Lienhart07} have been successfully tested in the case of rubidium.  Starting from a laser source at 1560 nm, the desired wavelength at 780 nm is obtained after frequency doubling the source. It allows to take advantage of the optoelectronic devices developed for the telecommunications industry in the 1530-1565 nm band.

Concerning potassium, laser sources are usually either Ti:Sapphire lasers, cooled laser diodes \cite{Goldwin02} or antireflection-coated laser diodes \cite{Nyman106}. In these cases, the repumping frequency is obtained either with a second laser or by frequency shifts through acousto-optic modulators (AOM), which increases the size of the optical setup and induces losses of power for the AOM solution. In this letter, we adapt the previous principle of SHG to get a 767 nm compact laser source for the cooling of potassium. Our setup can be used for the three isotopes of potassium, the  fermionic ($^{40}$K) and the two bosonic ($^{39}$K and  $^{41}$K) ones, which have been used in various experiments: the first Fermi sea \cite{DeMarco99}, molecular BEC \cite{Greiner03}, atomic interferometry \cite{Roati04,Fattori08}... 

Our source consists in a single laser diode at 1534 nm frequency doubled in a Periodically Poled MgO:SLN waveguide. This diode is current modulated to generate the two required frequencies for cooling. After amplification in a semiconductor tapered amplifier, this laser source has been employed to produce a Magneto-Optical Trap (MOT) of $^{39}$K. We get in this way a simple and compact setup for a fair price. Moreover, this apparatus is almost insensitive to vibrations and misalignments due to the use of fiber components.

\section{Experimental setup}

\begin{figure}[t!]
\centering
\rotatebox[origin=rB]{90}{
\includegraphics[width=5cm]{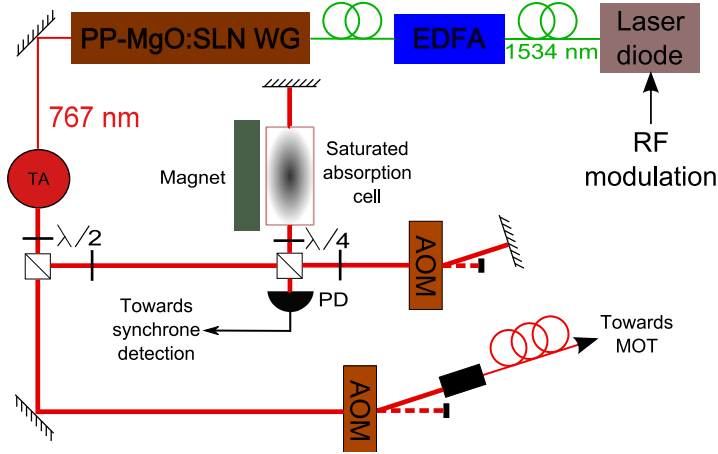}}
\caption{Scheme of the optical setup. A fibered distributed feedback 1534 nm  laser diode is current modulated at $\sim$ 462 MHz and amplified in an erbium-doped fiber amplifier (EDFA). After frequency doubling in an input-pigtailed waveguide (PP-MgO:SLN WG), the beam is amplified through a tapered amplifier (TA). While most of the light is sent to the science cell through an optical fiber to produce the MOT, a small fraction is used to lock the diode with a saturated absorption setup. The magnet is optional and can improve the saturated absorption signal in the presence of the sidebands (see text).}
\label{Table_optique}
\end{figure}

A scheme of the optical setup is presented on Fig. \ref{Table_optique}. A distributed feedback  1534 nm pigtailed laser diode (Fitel FOL 15DCWD-A81-19530-C, linewidth 1 MHz, 40 mW) is first current modulated with a radio-frequency to generate sidebands. One of them will be used to get the repumping frequency. The resulting beam is then amplified through a commercial 200 mW erbium-doped fiber amplifier (Keopsys) and frequency doubled in a Periodically Poled MgO:SLN input-pigtailed waveguide (HC Photonics Corp.). The chip is 32 mm long and 0.5 mm thick. We typically get $\sim$10 mW at 767 nm at the output. The input power is limited to 200 mW, the Epoxy glue used for the pigtail not being able to endure higher powers. It remains the most limitating aspect of this solution. The conversion efficiency would indeed be better, if we were able to inject more power in the waveguide. Recently, this problem has been solved by using others commercial waveguides \cite{Nishikawa09}. 

After the waveguide, the beam is amplified through a semiconductor tapered amplifier (EagleYard) to a power of 750 mW. After splitting, one part is used to lock the diode on the crossover  $^{39}$K transition through a saturated absorption setup after a double pass in a AOM (see Fig. \ref{Table_optique}). In this way, we can accurately detune the frequency from the excited hyperfine levels used for the cooling. The other part passes through a last AOM which acts as an optical switch. The light is finally injected into an optical fiber (the MOT fiber). It is connected to a 1-to-6 fiber beam-splitter (Sch$\rm{\ddot{a}}$fter und Kirchoff) which delivers six beams. We use them in a classical counterpropagating configuration to make a MOT of $^{39}$K. In a simple vacuum chamber, we typically get a few $10^8$ atoms loaded from potassium dispensers.

\section{Generation of the repumping frequency}

\begin{figure}[t!]
	\centering\resizebox{0.7\textwidth}{!}{
	%\rotatebox[origin=rB]{90}{
	\includegraphics[width = 5cm]{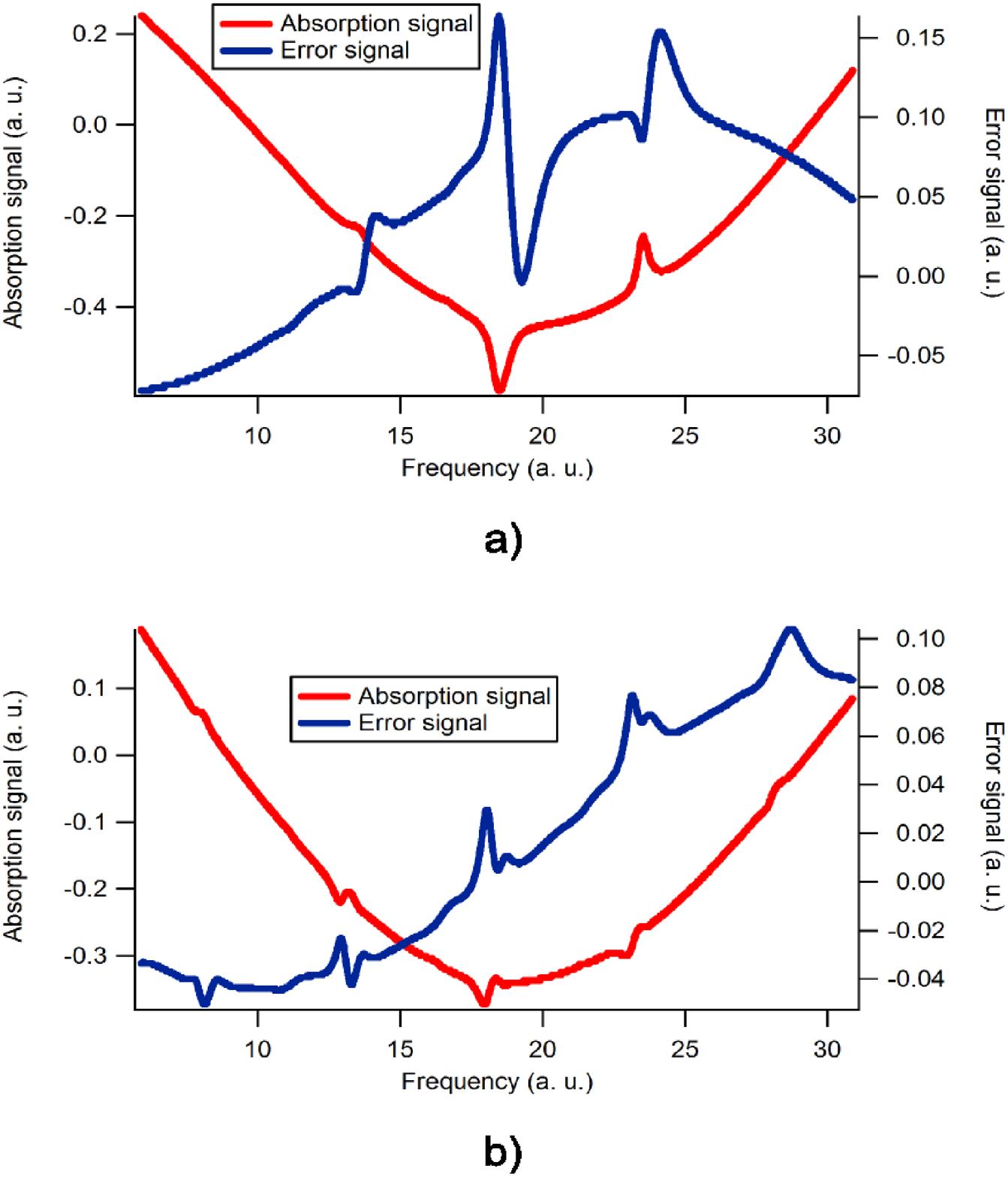}}%}
	\caption{Saturated absorption signal a) without current modulation b) with current modulation. In the second case, the error signal amplitude becomes smaller at the cross-over frequency. New dips appear due to the modulation, 462 MHz shifted from the initial ones. Adding a magnetic field on the absorption cell can increase the amplitude of the error signal.}	
		\label{Abs_sat}
\end{figure} 

Potassium requires a powerful repumping beam ($\sim$ 1/2) with respect to the cooling one compared with cesium or rubidium. Using current modulation rather than AOM for a powerful enough repumping frequency presents several advantages. It allows to reduce power losses, both cooling and repumping beams have the same polarization and optical mode, and the setup is much simpler. 

Our laser diode has a built-in radiofrequency current modulation bias tee. We apply a 462 MHz voltage on it, this frequency corresponding to the hyperfine splitting of the ground state. This generates two frequencies at $\pm 462$ MHz apart the carrier. When frequency doubled, it leads to one carrier, two sidebands at $\pm 462$ MHz and two others at  $\pm 924$ MHz with much less laser power. Only the carrier and the sideband at + 462 MHz will be used, others sidebands corresponding to power losses. We need a low RF power  ($\sim-3.4$ dBm) to get the right power balance, and contrary to \cite{Goldwin02}, the power repartition is stable. In front of the MOT fiber, we typically have 450 mW of light at 767 nm.

In absence of modulation (Fig. \ref{Abs_sat} a)), the saturated absorption dips $\rm 4^2S_{1/2}F=1\rightarrow 4^2P_{3/2}$ and $\rm 4^2S_{1/2}F=2\rightarrow 4^2P_{3/2}$ are visible, so is the crossover between the fundamental states. We lock the laser on the crossover peak.

Fig. \ref{Abs_sat} b) shows the saturated absorption signal affected by the modulation. When the modulation is applied, amplitudes of both cross-over and saturated absorption dips are reduced, and the error signal to lock the laser diode can become very poor. New absorption spectra due to the sidebands indeed overlap the initial one and mix with them. Adding a Zeeman shift with a magnet near the absorption cell ($\sim$50 G in our case) allows to rise degeneracy between pump and probe due to their opposite circular polarization; the signal can be improved in this way.

\section{Conclusion}

In conclusion, we have demonstrated a compact, simple and robust laser source to cool $^{39}$K by frequency doubling a laser diode at 1534 nm. The repumping frequency is generated by current modulation. In this way, the setup is simple and takes advantages of several fiber components at the telecoms wavelengths. Furthermore, power losses are reduced as compared with repumping frequency generation by an AOM. This device has been successfully employed to get a MOT of $^{39}$K. It could also work easily for $^{40}$K or $^{41}$K after minor changes \cite{Modulation}. The spectral linewidth at 767 nm can be narrowed down to a few kHz by using  an erbium-doped fiber laser for example instead of a laser diode.

Our setup will facilitate precision measurements applications of atom interferometry with  potassium atoms. Potassium offers the advantage compared with rubidium to allow the creation of Fermionic spin polarized samples or non interacting bosonic samples  with the help of Feshbach resonance \cite{Roati04,Fattori08}.   As in \cite{Stern09}, this setup could be adapted to cool the atoms and to coherently manipulate them with Raman transitions with the same device. A single beam with two frequencies for Raman transitions has the advantage to be insensitive to relative misalignment. We could also think about a mobile double species atomic interferometer to test the Universality of Free Fall \cite{Nyman06}, whose laser sources would be provided by SHG of telecoms wavelengths.

\section*{Acknowledgments}

This research was supported by CNRS and CNES as part of the ICE project and the Locabec and Miniatom projects from ANR. This work is also funded by the European Space Agency under the SAI program and  by the European Science Foundation (EUROQUASAR project). Further support comes from the European Union STREP consortium FINAQS. BA salary is funded by D$\rm\acute{e}$l$\rm\acute{e}$gation G$\rm\acute{e}$n$\rm\acute{e}$rale de l'Armement and BB by RTRA
"Triangle de la physique". Laboratoire Charles Fabry is member of IFRAF \cite{Ifraf}.


\begin{thebibliography}
\bibitem{Modulation}	For $^{41}$K, the hyperfine splitting of the ground state is 254 MHz, and current modulation is still possible. For $^{40}$K, the hyperfine splitting is 1.28 GHz, larger than the bandwidth of the modulation bias tee. However, we were able to generate two sidebands at $\pm642$ MHz, with modulation amplitude high enough to extinguish the carrier wave. Higher order sidebands are still generated, which result in power losses.
\end{thebibliography}
\end{document}